# Two-dimensional electron gas in monolayer InN quantum wells

W. Pan[1,a)], E. Dimakis[2,a), b)], G.T. Wang[1], T.D. Moustakas[2], and D.C. Tsui[3]

[1] Sandia National Laboratories, Albuquerque, New Mexico 87185, USA

[2] Boston University, Boston, Massachusetts 02215, USA

[3] Princeton University, Princeton, New Jersey 08544, USA

ABSTRACT

We report in this letter experimental results that confirm the two-dimensional nature of the electron systems in monolayer InN quantum wells embedded in GaN barriers. The electron density and mobility of the two-dimensional electron system (2DES) in these InN quantum wells are $5 \times 10^{15}$ cm$^{-2}$ and 420 cm$^2$/Vs, respectively. Moreover, the diagonal resistance of the 2DES shows virtually no temperature dependence in a wide temperature range, indicating the topological nature of the 2DES.

a) Authors to whom correspondence should be addressed. Electronic addresses: wpan@sandia.gov and e.dimakis@hzdr.de
b) Current address: Helmholtz-Zentrum Dresden-Rossendorf, Institute of Ion Beam Physics and Materials Research, 01328 Dresden, Germany.

InN, a relatively new comer in the family of III-nitride (III-N) semiconductors, has attracted a great deal of current interest. Indeed, with a bulk band gap of ~ 0.7eV[1,2], InN possesses the narrowest band gap within the III-N family. As a consequence, integration of InN with other III-N materials can greatly enlarge the operational spectrum from the infrared to deep ultra-violet, thus presenting unprecedented potential for applications in light-emitting diodes, lasers, photovoltaics, sensors, etc.[3,4]

Very recently, InN has also gained attention as a novel electronic material for exploring condensed matter topological physics. It was recently predicted[5] that, if the InN layer thickness is larger than 4 monolayers for an InN quantum well embedded between GaN barriers, its band gap can turn negative, i.e., the bottom of conduction band is lower than the top of valence band. This inverted band structure, combined with a relatively strong spin-orbit coupling in InN, can render the InN quantum well a new two-dimensional topological insulator. Compared to other two-dimensional topological insulator systems, such as HgTe[6,7] and InAs/GaSb[8,9], the inherent transparence of InN quantum well systems could bring new applications in opto-topological electronic devices.

It is well known that, unlike other III-N materials, the growth of InN quantum well is extremely difficult. First, due to a high vapor pressure of indium (In), the growth temperature is limited to a much lower value than other III-N materials[10]. Moreover, indium tends to form clusters during the growth. Recently, a new growth technique was developed to grow single and/or bilayer quantum well structures embedded in GaN barriers[11,12]. Detailed TEM analysis and subsequently photoluminescence and electro-luminescence studies confirmed a high fidelity in InN quantum well growth.

Compared to relatively extensive structural and optical characterizations, there are few experimental studies on the transport properties of the electron system realized in InN quantum wells. In fact, to our knowledge, no experimental evidence has ever been obtained confirming the two-dimensional nature of the electron system in InN quantum wells.

In this letter, we present experimental results from our electron transport studies confirming that indeed the electron system in InN quantum wells is of two-dimensional nature. Furthermore, the diagonal resistance $R_{xx}$ of the two-dimensional electron system (2DES) shows virtually no temperature dependence from 0.3 to ~ 20K. We propose that this temperature independent behavior may be related to the topological nature of the 2DES in our InN quantum wells, as predicted in Ref. [5].

The monolayer (ML) InN quantum wells were grown on a c-plane sapphire substrate by radio frequency plasma-assisted molecular beam epitaxy. After a thick (~ 4 µm) GaN buffer layer, a superlattice structure of 40 InN quantum wells consisting of 1ML of InN embedded between 10 nm GaN barriers and a 50nm-thick GaN cap layer at the top were grown. The growth was carried out under In-rich conditions at ~700°C and this temperature was kept the same for both the InN QWs and GaN barriers[12].

Standard high-resolution cross-section transmission electron microscopy was used to show that the thickness of InN quantum well is indeed 1ML[12]. The InN monolayers were grown pseudomorphically on the GaN barriers with no misfit networks observed at the InN/GaN interfaces. The ultrathin InN layer is coherently strained with the GaN barrier layer with the interface being smooth and abrupt. The InN layers appear to be continuous, with uniform thickness.

Figure 1 shows the cathodoluminescence (CL) results carried out on a freshly cleaved cross-section. A strong CL peak is seen at ~ 400nm or 3.1eV, consistent with previous photoluminescence results[12]. This CL peak energy is much larger than the InN bulk band gap of ~ 0.7eV, and is believed to be due to the quantum confinement effect[12,13]. The second peak at ~ 800nm is probably the second harmonic of the 400nm peak. No other discernible peaks are observed in the wavelength regime between 350 and 875nm.

In the following we will concentrate on the electron transport results obtained in a monolayer InN quantum well structure that was cut from the same wafer as in Figure 1. Pure indium, annealed at 440°C in a forming gas environment, was used to achieve ohmic contacts to the electron system in the InN quantum wells. Conventional low-frequency (~ 11Hz) lock-in technique was used to measure the diagonal resistance $R_{xx}$ and Hall resistance $R_{xy}$ at low temperatures and in tilt magnetic fields.

The main result of this paper is shown in Figure 2, where the Hall resistance $R_{xy}$ is plotted as a function of perpendicular magnetic field ($B_{perp}$) and $B_{perp} = B_{total} \times \cos(\theta)$ at a sample temperature of ~ 15K. Here, $B_{total}$ is the total external magnetic field and $\theta$ the tilt angle. It is clearly seen that $R_{xy}$ at two angles, $\theta = 0$ and $60^o$, overlaps with each other perfectly. We emphasize that the overlap in $R_{xy}$ between $\theta = 0$ and $60^o$, when plotted versus the perpendicular magnetic field, confirms that the electron system in the InN quantum wells is of two dimensional nature [14].

From the Hall coefficient around B = 0, the electron density is estimated as $5 \times 10^{15}$ cm$^{-2}$. This translates to a density of $1.25 \times 10^{14}$ cm$^{-2}$ for each quantum well, assuming the electron density is the same in all quantum wells. This density, together with the zero field resistivity allows us to deduce the electron mobility as ~ 420 cm$^2$/Vs. We also note here that it was observed that the electron density and mobility varied form one cool-down to another, within 20%.

Figure 3 shows the resistivity $\rho_{xx}$ as a function of magnetic field at two selected temperatures of 1.5 and 0.3K. Here, $\rho_{xx}$ is obtained from $R_{xx}$ after taking into account the geometrical ratio. Two features are striking. First, $\rho_{xx}$ shows no field dependence up to $\pm$ 7.5T. This is unusual as non-zero magnetoresistance is a rather common phenomenon in disordered 2DES due to either quantum mechanical or classical origins. Quantum mechanically, electron-electron interaction, spin-orbit coupling, or quantum interference effect can induce either positive or negative magnetoresistance[14]. Classically, magnetoresistance generally follows the well-known Kohler's rule[15] and $\rho_{xx} \propto 1 + (\mu B)^2$. Using the obtained mobility, a 10% increase in $\rho_{xx}$ is expected at 8T, which was not observed. Finally, we want to point out that a vanishing magnetoresistance (VMR) was observed in the past in polyacetylene nanofibers under high electric fields[16]. This VMR was believed to be due to the deconfinement of spinless charged soliton pairs in high electric fields. It would be interesting to explore in the future whether the absence of magnetoresistance in our InN quantum wells is related to the strong polarization field as we are going to discuss below.

Second, $\rho_{xx}$ shows virtually no temperature dependence. In fact, as shown in the inset of Figure 3, $\rho_{xx}$ is temperature independent from 0.3 to ~ 20 K. First, we emphasize that a temperature independent behavior of $\rho_{xx}$ over such a large range of temperature is not expected. Usually, in a high density disordered 2DES, weak localization effect is expected to dominate the temperature dependence of $\rho_{xx}$ at low temperatures[14]. On the other hand, $K_F l >> 1$ in our InN quantum wells due to a very high electron density of the 2DES, where $K_F$ is the Fermi wavevector and $l$ the elastic mean free path. Consequently, the 2DES is deep in the metallic regime and no metal-insulator transition is expected as the sample temperature is lowered[17]. There are conventional mechanisms that can lead to a weak temperature dependent or independent behavior of $\rho_{xx}$. For example, the electron mobility in InN quantum well is limited by elastic scattering mechanisms – alloy scattering, interface roughness scattering, or both[18]. While we cannot eliminate these mechanisms for the observed temperature independent behavior of $\rho_{xx}$, yet, a constant $\rho_{xx}$ over a temperature range of nearly two orders of magnitude is strikinging. In this regard, we want to point out another intriguing possibility, i.e., this temperature independent behavior of $\rho_{xx}$ may be related to the InN quantum wells being a two-dimensional topological insulator[5]. Indeed, temperature independent behavior of $\rho_{xx}$ has been observed in a two-dimensional topological insulator HgTe[19], and a three-dimensional topological insulator SmB$_6$[20]. There, the absence of temperature dependence is believed to be due to the edge states (in 2D cases) and surface states (in 3D cases), which are robust and protected by time-reversal symmetry. It is possible that the observed temperature independent $\rho_{xx}$ in our InN quantum wells is also due to our InN quantum wells being a two-dimensional topological insulator. It is known that in ultra-thin InN quantum wells, due to a ~ 11% lattice-constant mismatch between InN and GaN, a large strain field exists in the InN QWs[21]. The strain field then produces large polarization charges and, thus, a large electric field perpendicular to the InN QWs. This polarization field then leads to localizations of hole and electron states on the opposite sides of the QWs and bring the electron and hole states closer in energy[5]. Consequently, the energy gap is reduced. Furthermore, the difference in the spontaneous piezoelectric polarization between InN and GaN is expected to enhance this reduction in the energy gap and may even induce an energy band inversion[5,22]. This inverted band gap, along with a relatively strong spin-orbit coupling in InN QWs, can drive the InN quantum wells from a conventional semiconductor to a two-dimensional topological insulator, as predicted in Ref. [5]. The edge states in this two-dimensional topological insulator then can give rise to a temperature independent behavior of $\rho_{xx}$ as observed in other topological insulators[19,20].

In summary, we have demonstrated through our electron transport studies that the electron system realized in the monolayer InN quantum wells are of two-dimensional nature. We also propose that the observed temperature independent behavior in the diagonal resistance $R_{xx}$ may be related to the 2DES in InN quantum wells being in the two-dimension topological insulator regime.

We thank X. Shi for experimental assistance and J.L. Reno for helpful discussion. The work at Sandia and Princeton University was supported by the Department of Energy, Office of Basic Energy Sciences, Division of Materials Sciences and Engineering. Sandia National Laboratories is a multi-program laboratory managed and operated by Sandia Corporation, a wholly owned subsidiary of Lockheed Martin Corporation, for the U.S. Department of Energy's National Nuclear Security Administration under contract DE-AC04-94AL85000. The work at Boston University was supported by the Department of Energy under Grant No. DE-FG02-06ER46332.

**Figures and Figure Captions**

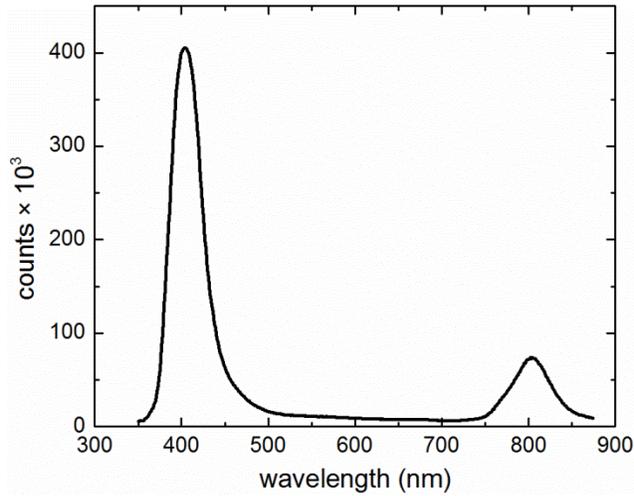

**Figure 1**. Room temperature cathodoluminescence of our 10 monolayer InN quantum wells on a freshly cleaved cross-section. The main peak at ~ 400 nm is due to the InN quantum wells. The weaker peak at ~ 800 nm is probably the second harmonics of the 400nm peak.

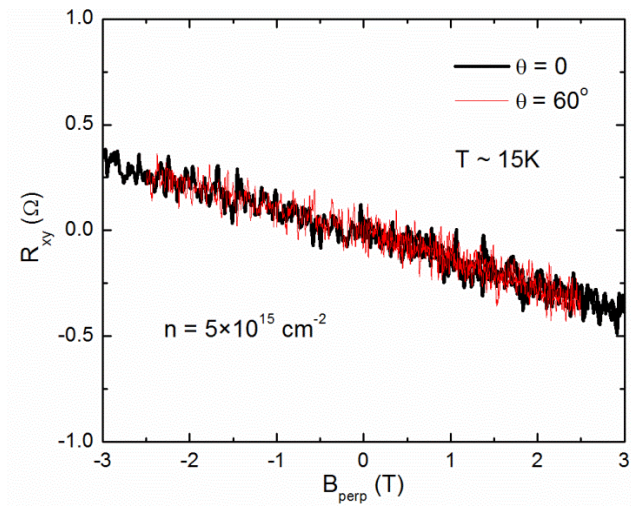

**Figure 2**. Hall resistance $R_{xy}$ versus perpendicular magnetic field. Two curves at zero tilt and 60 degrees are shown. They overlap with each other.

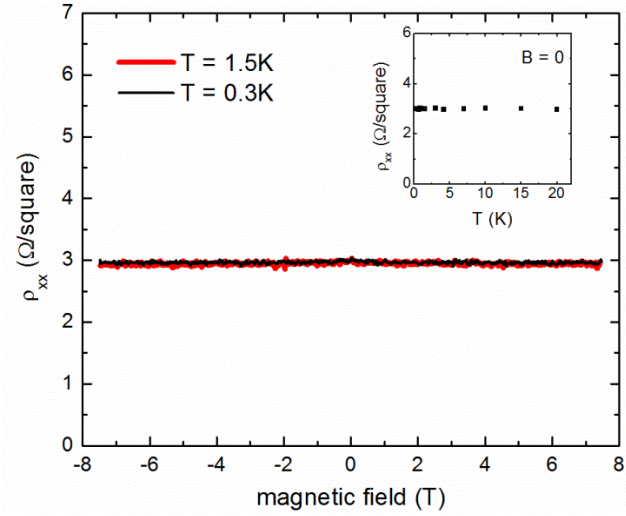

**Figure 3**. $\rho_{xx}$ as a function of magnetic field at two sample temperatures of 1.5 and 0.3 K. $\rho_{xx}$ shows virtually no magnetic field and temperature dependence (also shown in the inset).